\documentclass{emulateapj}

\usepackage{graphicx}
\usepackage{epstopdf}
\usepackage{color}

\def\gr{$\gamma$-ray}

\shorttitle{VHE gamma-ray backgorund}
\shortauthors{Neronov \& Semikoz}

\begin{document}

\title{Extragalactic Very-High-Energy gamma-ray background}
\author{A.~Neronov}
\affil{ISDC Data Center for Astrophysics, Chemin d''Ecogia 16, 1290 Versoix,
Switzerland and Geneva Observatory, 51 ch. des Maillettes, CH-1290 Sauverny,
Switzerland}
\and
\author{D.V.~Semikoz}
\affil{APC, 10 rue Alice Domon et Leonie Duquet, F-75205 Paris Cedex 13, France}
\affil{Institute for Nuclear Research RAS, 60th October Anniversary prosp. 7a,
Moscow, 117312, Russia}

\begin{abstract}
We study the origin of the extragalactic diffuse gamma-ray background using the data from the {\it Fermi} telescope. To estimate the background level, we count photons at high Galactic latitudes $|b|>60^\circ$. Subtracting photons associated to known sources and the residual cosmic ray and Galactic diffuse backgrounds,  we  estimate the Extragalactic Gamma-ray Background (EGB) flux.  We find that  the spectrum of EGB in the very-high-energy (VHE) band above 30~GeV follows the stacked spectrum of BL Lacs.  LAT data reveal the positive $(1+z)^k$, $1<k<4$ cosmological evolution of the BL Lac source population  consistent with that of their parent population, FR I radio galaxies. We show that EGB at $E>30$~GeV could be completely explained by emission from unresolved BL Lacs if $k\simeq 3$. 
\end{abstract}

\section{Introduction}

Study of diffuse background emission produced by faint sources with flux levels below the sensitivity of a telescope is commonly used to constrain the nature of source populations in the Universe and their cosmological evolution. In the high-energy \gr\ band (HE, 0.1-100 GeV) the diffuse extragalactic \gr\ background (EGB) was detected for the first time by {\it SAS-2} satellite \citep{fichtel78}, was further studied by EGRET telescope on board of {\it CGRO} mission \citep{sreekumar98,strong04} and, most recently, by the Large Area Telescope (LAT) on board of {\it Fermi} satellite \citep{fermi_background}. 

It is often assumed that the dominant contribution to the EGB is given by distant Active Galactic Nuclei (AGN), in particular, blazars \citep{padovani93,stecker93,chiang95,stecker96,mukherjee99,mucke00,inoue09}. However, a recent study by {\it Fermi} collaboration reveals that blazars might contribute only a relatively small fraction of the HE EGB level \citep{fermi_agn}, while a significant part of the EGB should be either explained by a yet unknown source population or have a truly diffuse nature (see, however, \citet{stecker10}).

The EGB in the Very-High-Energy (VHE, \gr s in the $E\gtrsim 100$~GeV) range has never been measured. On one hand, the effective collection area of previous space-based \gr\ telescopes was not sufficient to achieve significant photon statistics in this energy band. On the other hand, the efficiency of cosmic ray background rejection in the ground-based Cherenkov \gr\ telescopes, like HESS, MAGIC and VERITAS is not sufficient for detection of the isotropic diffuse EGB on top of the cosmic ray background. Thus, the properties and the origin of VHE EGB remain largely unconstrained up to now.

It is clear that the VHE EGB should contain a contribution from the unresolved point sources. The main candidate source class is, as in the case of HE EGB, that of blazars. At the same time,  the VHE EGB could contain, apart from the contribution from unresolved extragalactic point sources, genuine diffuse components which could be produced via several mechanisms.  For example, if the spectra of a large number of  \gr -loud AGNs extend to the energies above 300~GeV, all the power emitted initially in 
\gr s with energies higher than $\sim 300$~GeV  is absorbed in the pair
production of \gr s on the cosmological infrared and/or  microwave backgrounds \citep{gould67}.
Secondary inverse Compton emission of electron-positron pairs deposited in the
intergalactic space in result of the pair production leads to generation of
diffuse extragalactic emission in the VHE energy band \citep{coppi97}.  Another mechanism
which can lead to the generation of diffuse component of VHE EGB  is electromagnetic cascade initiated in the intergalactic space by ultra-high energy cosmic rays (UHECR) interacting with cosmic microwave background
photons \citep{berezinsky75,semikoz09,berezinsky11}. The cascade channels the power from the highest energies of about
$10^{20}$~eV down to the $\sim 0.1$~GeV  band in which the mean free path of the
\gr s becomes comparable to the size of the visible part of the Universe. 

Apart from the "guaranteed" (but, possibly, very weak) diffuse contributions, isotropic VHE \gr\ background might contain contributions from "exotic" diffuse sources, like diffuse emission from annihilation of Dark Matter particles in the outer halo of the Milky Way galaxy and the annihilation signal accumulated from the dark matter halos of all galaxies in the course of cosmological evolution \citep{fermi_DM}.

Whatever are the sources of VHE EGB, they are scattered across the Universe, so that a significant contribution to the flux is produced at redshifts $z\sim 1$. A known effect of absorption of VHE \gr s due to the interactions with infrared/optical Extragalactic Background Light (EBL) should lead to attenuation of the $E>50$~GeV signal produced by the sources at large redshifts $z\sim 1$ \citep{gould67,kneiske04,franceschini08,stecker_ebl,gilmore09}. This should leave an "imprint" on the VHE EBL spectrum, which should have the form of a gradual suppression with the increasing photon energy. Detecting a EBL suppression feature in the EGB spectrum would provide an important constraint on the (largely uncertain) evolution of the EBL density and spectrum up to redshifts $z\sim 1$. Such a constraint is otherwise difficult to obtain from the studies of individual extragalactic VHE \gr\ sources because of the limited signal statistics at the highest  energies, especially for the sources at significant redshifts.

In what follows we discuss the measurement of  the EGB  derived from the data of  {\it Fermi}/LAT telescope \citep{fermi_description}. The measurement is obtained from the counting of  photons at high Galactic latitudes, after subtraction of the Galactic diffuse emission and the residual cosmic-ray background not rejected by the LAT data analysis software. We compare the measurement of EGB obtained in this way with the measurement previously derived from the likelihood analysis of all-sky data by \citet{fermi_background}.

The EGB flux above 30~GeV turns out to be comparable to the flux in extragalactic VHE \gr\ sources resolved by {\it Fermi}. We find that the spectrum of EGB in this energy range follows the cumulative spectrum of the resolved sources. 
Dominant population of extragalactic VHE \gr\ sources is BL Lacs. Noticing the similarity of the spectrum of VHE EGB and of the cumulative BL Lac VHE \gr\ spectrum, we put forward a hypothesis that the VHE EBL is produced by unresolved BL Lacs with fluxes below the sensitivity of LAT. We explore this hypothesis and show that it could be valid if BL Lacs follow a positive cosmological evolution pattern, characteristic for other types of AGN, in particular for the parent population of BL Lacs objects, Fanaroff-Riley type I (FR I) radio galaxies.

\section{Data selection and data analysis} 

For our analysis we consider all publicly available LAT data from August 4, 2008 to January 23, 2011. We process the data using {\it Fermi} Science Tools\footnote{\tt http://fermi.gsfc.nasa.gov/ssc/data/analysis/scitools/}. We filter the entire data set with {\it gtselect} and {\it gtmktime} tools following the recommendations of {\it Fermi} team\footnote{\tt http://fermi.gsfc.nasa.gov/ssc/data/analysis/scitools/} and  retain only events belonging to "ultraclean" ({\tt P7ULTRACLEAN\_V6}) event class, which has minimal residual cosmic ray contamination.  

Estimate of the contribution of point sources to the total flux requires separation of the photons coming from the point sources from those produced by the diffuse emission. Such separation is most straightforward for the photons with narrow point-spread-function (PSF). Taking this into account, select two sub-classes (which provide dominant contribution to the ultraclean events) with the most compact PSF, the sub-classes selected by imposing the selection criterium {\tt EVENT\_CLASS=65311} or {\tt 32543}. Other sub-classes of the ultraclean events have worse PSF. Point source contribution in these photons  suffers from  an additional uncertainty. Taking this into account, we restrict our attention to the subset of the ultraclean events with the best PSF.   

We retain events with Earth zenith angle $\theta_z\le 100^\circ$. To estimate the flux from the photon counts we use {\it gtexposure} tool.  We consider only events at high Galactic latitudes, in the regions $|b|\ge 60^\circ$. 

Our analysis is based on the so-called "Pass 7" selection of the LAT data (see {\tt http://fermi.gsfc.nasa.gov/ssc/data/access/}). However, use use a comparison of the Pass 7 data with the previous Pass 6 data selection in the estimate residual cosmic ray contamination of the set of events chosen for the analysis. The residual cosmic ray fraction in the Pass 6 data was studied in details by \citet{fermi_background}. Re-calculation of the residual cosmic ray fraction for any new selection of events, including the one considered in our analysis, could be done in a straightforward way as explained in Section \ref{sec:cr}.

\section{Diffuse $\gamma$-ray background}

Signal detected by LAT at high Galactic latitudes contains four types of contributions: emission from point sources, diffuse \gr\ emission from the Galaxy, EGB and residual cosmic ray background not rejected by the analysis software. To measure the EGB flux, one needs to separate the contributions from the four components in the overall signal in a given energy band. 

\subsection{Point source contribution}
\label{sec:ps}

\begin{figure}
\includegraphics[height=\linewidth,angle=-90]{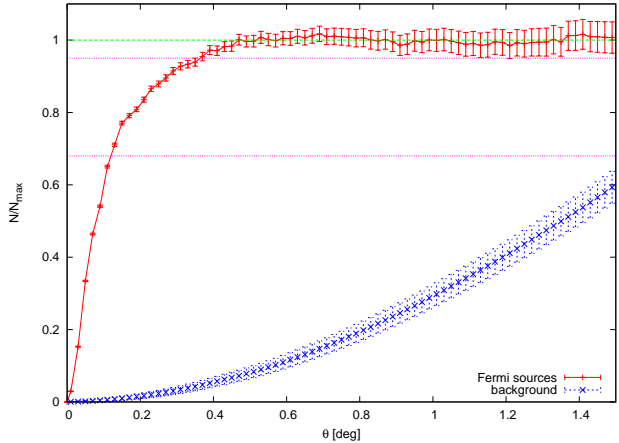}
\caption{Cumulative front photon distributions around point sources in 12.5-25 GeV energy band. Red points show the source photons, blue points show the background. Horizontal lines (from top to bottom) show the 100\%, 95\% and 68\% levels.}
\label{fig:PSF_cumulative}
\end{figure}

Point source component could be singled out in a straightforward way if the set of sources detectable in a given energy band is known. To define the set of sources we find the sources correlating with the arrival directions of photons in each energy band, using the method described by \citet{100GeV_sky}\footnote{See {\tt http:/www.isdc.unige.ch/vhe/index.html} for an updated version of the VHE source list.}. To calculate the total number of photons associated to the sources, we construct a cumulative distribution of photons as a function of the distance $\theta$ from the source and split it on the background and source contributions. The background contribution grows asymptotically as $\theta^2$, while the source contribution asymptotically reaches constant. An example of the cumulative photon distribution around the source positions in the 12.5-25 GeV energy band is shown in Fig. \ref{fig:PSF_cumulative}.

\subsection{Galactic diffuse emission contribution}
\label{sec:gal}

Contribution of the diffuse emission from the Galaxy should be found via a detailed fitting the all-sky photon distribution to an all-sky spatial and spectral template. This contribution is best constrained by the all-sky photon distribution in the 0.1-10~GeV energy band, where event statistics is very high. Detailed fitting of the Galactic diffuse emission to the data in the 0.1-10~GeV band was done by \citet{fermi_background}. In our analysis we rely on the best-fit model of Galactic diffuse emission derived by \citet{fermi_background}. This model is available in the sky region of interest, $|b|\ge 60^\circ$, see Fig. 6 of the Supplemental Material in the Ref. \citet{fermi_background}. The uncertainties of the Galactic diffuse emission model are also discussed by \citet{fermi_background}. We take these uncertainties into account.

\begin{figure}
\includegraphics[width=\linewidth]{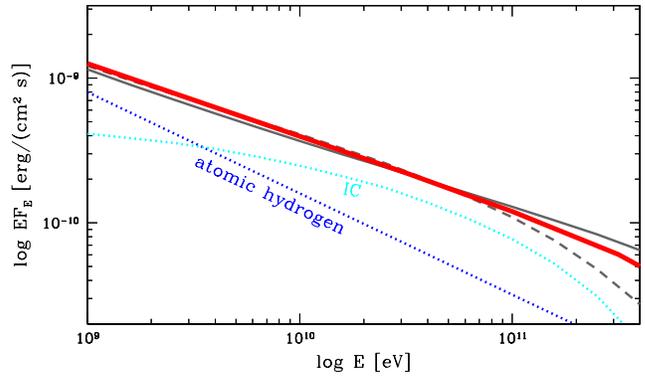}
\caption{Extrapolation of the spectrum of the Galactic diffuse emission at high Galactic latitudes $|b|\ge 60^\circ$ to the  100~GeV energy range. Blue and cyan dotted lines below 100~GeV show the contributions from cosmic ray interactions with interstellar medium and inverse Compton scattering from cosmic electrons calculated by \citet{fermi_background}. Continuation of the cyan dotted line above 100~GeV is calculated assuming that inverse Compton emission is produced by electrons with a cut-off powerlaw spectrum with cut-off at 1~TeV. Grey dashed line is the sum of the cosmic ray and inverse Compton contributions. Solid grey line shows the overall diffuse emission spectrum in which the inverse Compton emission is produced by electron distribution without high-energy cut-off at 1~TeV. Red thick solid line shows the spectrum used for subtraction of Galactic component from the overall high Galactic latitude diffuse emission flux. }
\label{fig:Galactic}
\end{figure}

The model consists of two main contributions: the "atomic hydrogen" component produced by interactions of cosmic rays with interstellar matter and "inverse Compton"  (IC) component produced by inverse Compton emission from cosmic ray electrons. Extrapolation of the atomic hydrogen component to the highest LAT energies is straightforward: the pion decay spectrum follows the cosmic ray spectrum and extrapolation has the form of a simple powerlaw with photon index $\sim 2.7$, the same as the slope of the cosmic ray spectrum.  This component gives a sub-dominant contribution above 100~GeV. Extrapolation of the IC component depends on the unknown shape of the (average over interstellar medium) cosmic ray electron spectrum at the energies above TeV. We have checked that in the model of \citet{fermi_background} the spectrum of IC component is consistent with the spectrum of IC scattering of the local interstellar radiation field \citep{moskalenko06} by electrons with the spectrum $dN_e/dE\sim E^{-3}\exp\left(-E/1\mbox{ TeV}\right)$. This electron spectrum consistent with the cosmic ray electron spectrum observed on the Earth \citep{fermi_electrons,HESS_electrons}.  The IC spectrum produced by such electron population is shown by the cyan dotted line in Fig. \ref{fig:Galactic}. The overall Galactic diffuse emission spectrum at high Galactic latitudes is then the sum of the atomic hydrogen and IC contributions, shown by the dashed black line in Fig. \ref{fig:Galactic}.

The high-energy cut-off of the local cosmic ray electron spectrum is most likely determined by the distance to the closest cosmic ray electron sources, e.g. to the closest pulsar wind nebulae \citep{aharonian04}, rather than by the intrinsic cut-off in the injection spectrum of electrons from the sources. This means that the local measurement of the high-energy cut-off of the cosmic electron spectrum does not provide a measurement of the high-energy cut-off in the injection spectrum of electrons. It is possible that Galactic cosmic electron sources inject electrons with energies much higher than $\sim 1$~TeV. This possibility is shown by the solid grey line in Fig. \ref{fig:Galactic} which shows the sum of the atomic hydrogen contribution with the IC emission from electrons without high-energy cut-off in the spectrum. The IC component still exhibits suppression at the energies $\sim 1$~TeV because of the Klein-Nishina effect.

To take into account the above mentioned uncertainty of the IC component we adopt an approximation $dN_\gamma/dE\sim E^{-2.5}\exp\left(-E/2\mbox{ TeV}\right)$ for the high Galactic latitude diffuse emission spectrum at $E>100$~GeV. This approximation is shown by the red thick solid line in Fig. \ref{fig:Galactic}. This spectrum lies exactly in the middle between the two extreme possibilities: TeV-scale high-energy cut-off in the cosmic electron spectrum and no cut-off in the cosmic electron spectrum. One should take into account that the uncertainty of this approximation reaches $\simeq 50$\% at the highest energies. We take this uncertainty into account in the calculation of the EGB spectrum, by adding it as a systematic error. The two extreme possibilities for the behavior of electron spectrum above 1~TeV (exponential cut-off exactly at 1~TeV and no cut-off at all) provide a good estimate of the overall uncertainty of electron spectrum in the interstellar medium in this range. The uncertainty of the shape of electron spectrum dominates the uncertainty of the inverse Compton component of Galactic diffuse emission at high Galactic latitudes.

\subsection{Residual cosmic ray background contribution}
\label{sec:cr}

To estimate the residual cosmic ray background in the set of events selected for the analysis, we rely on the knowledge of residual 
The residual cosmic ray background in the {\tt dataclean} event class of Pass 6 data is extensively discussed in \citet{fermi_background}. The residual cosmic ray background in the subset of Pass 7 {\tt superclean} events used in our analysis could be calculated from the known residual cosmic ray background in the Pass 6 {\tt dataclean} events via a straightforard comparison of statistics of events on- and off-point-sources in the two classes.

First, the residual cosmic ray fraction in the Pass 6 {\tt dataclean} events should be calculated from the known suppression factor of cosmic ray event at transition from the {\tt diffuse} event class to the {\tt dataclean} event class in Pass 6 (see \citet{fermi_background} for the detailed discussion of the suppression factor). 
In each of the two event classes, the entire event set consists of a certain number of \gr\ events $N_{\gamma, i}$ and a certain number of residual cosmic ray events, $N_{CR, i}$ where $i$ stands for ${3+4}$ or $4$. Cleaning of the event set done to produce the {\tt dataclean} event set from {\tt diffuse} set results in rejection of a large fraction of the cosmic ray events, $N_{CR,4}=\alpha_{CR} N_{CR,3+4}$ with $\alpha_{CR}\ll 1$. However, it results also in rejection of a number of true \gr\ events, so that $N_{\gamma,4}=\alpha_\gamma N_{\gamma,3+4}$ with $\alpha_\gamma<1$. 

The suppression factor $\alpha_{CR}$ is known as a function of energy from the Monte-Carlo simulations of cosmic ray and \gr\ induced events in the LAT detector by \citet{fermi_background}. The suppression factor $\alpha_\gamma$ could be found directly from the data set, by comparing statistics of events coming from the point sources in the {\tt diffuse} and {\tt dataclean} event classes (see section  \ref{sec:ps} above). In the calculation of $\alpha_\gamma$ all the photons associated to $\sim 10^3$ point sources listed in the Fermi 2-year catalog \citep{fermi_catalog} could be used. This provides very large event statistics so that uncertainty of $\alpha_\gamma$ is negligible. Knowing the total numbers of events in the two event classes $N_{tot,i}$ one can resolve the system of equations 
\begin{equation}
\left\{
\begin{array}{l}
N_{CR,4}/\alpha_{CR}+N_{\gamma,4}/\alpha_{\gamma}=N_{tot,3+4}\\
N_{CR,4}+N_{\gamma,4}=N_{tot,4}
\end{array}
\right.
\end{equation}
with respect to $N_{\gamma,4}, N_{CR,4}$ to find the residual cosmic ray background in each energy bin  for the {\tt dataclean} event class. 

The residual cosmic ray fraction in the sub-class of the Pass 7 {\tt superclean} events used in our analysis is then estimated in a similar way, once the residual cosmic ray fraction $\kappa_4$ in the point-source-subtracted set of the Pass 6 events, $N_{CR,\rm off,4}=\kappa_{4}N_{\rm off,4}$, is known.

Indeed, the transition from the Pass 6 {\tt dataclean} events to the Pass 7 events belonging to the event classes  65311 and 32543 leaves a fraction $\alpha_{\gamma,6\rightarrow 7}$ of \gr\ events (actually, $\alpha_{\gamma, 6\rightarrow 7}>1$ in a broad energy range around 10 GeV). It also suppresses or increases the residual cosmic ray background, so that the residual cosmic ray fraction in the off-source events changes from $\kappa_4$ to $\kappa_{7}$. The off-source events in the two classes are then the sum  of the diffuse \gr\ emission photons and of the residual cosmic rays:
\begin{equation}
\left\{
\begin{array}{l}
\kappa_4 N_{\rm off,4}+N_{\gamma,\rm off, 4}=N_{\rm off,4}\\
\kappa_{7} N_{\rm off, 7}+\alpha_{\gamma, 6\rightarrow 7}N_{\gamma,\rm off, 4}=N_{\rm off,7}\\
\end{array}
\right.
\end{equation}
Knowing the statistics of the off-source events in the Pass 6 and Pass 7 events, $N_{\rm off,4}$ and $N_{\rm off,7}$, one could find the residual cosmic ray fraction in the Pass 7 data 
\begin{equation}
\kappa_7=1-\alpha_{\gamma,6\rightarrow 7}(1-\kappa_6)\frac{N_{\rm off,4}}{N_{\rm off,7}}
\end{equation}
The resulting estimates of the level of residual cosmic ray background for  the events selected in the Pass 7 data in energy bins between 3 and 100~GeV are shown by the grey data points in Fig. \ref{fig:spectrum}. 

From Fig. \ref{fig:spectrum} one could see that the contribution of the residual cosmic rays to the signal at 100~GeV is likely to be small. However, extrapolation of the estimate of efficiency of rejection of the residual cosmic ray background much above 100~GeV  is highly uncertain. It is possible that the efficiency of rejection of both the nuclear and electron/positron component of the cosmic ray flux drops because of the similarity of the cosmic ray and $e^+e^-$ pair tracks with large Lorentz factors. Inefficient rejection of the residual cosmic rays might lead to the contaminate the diffuse background signal and lead to a large over-estimation of the diffuse background flux. Because of this problem, we are able to only derive an upper limit on the EGB at the energies much above 100~GeV (for the energy band at 100~GeV we show a comparison between the 95\% confidence level upper limit and the measurement). A proper measurement of the EGB flux at the highest energies accessible to LAT would require extensive Monte-Carlo simulations taking into account detector response \citep{ackerman_texas}.

\subsection{Extragalactic \gr\ background spectrum}
\label{sec:spectrum}

\begin{figure}
\includegraphics[height=\columnwidth]{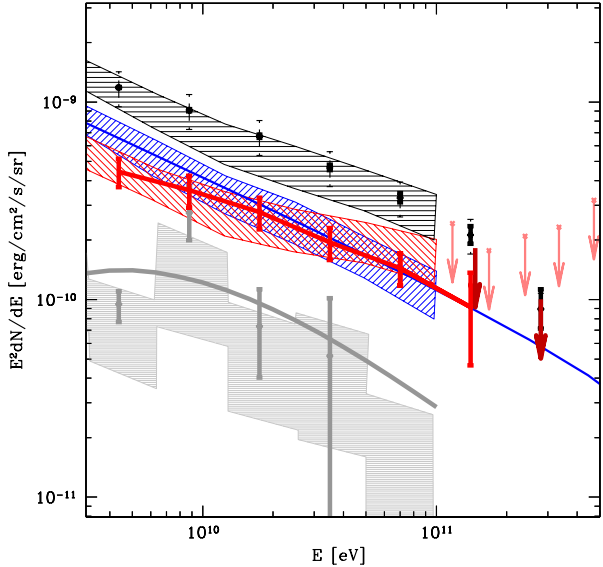}
\caption{Estimate of the flux of isotropic component of diffuse emission obtained by the direct photon counting method (red thick line, data points and upper limits). For comparison, the spectrum of isotropic component of diffuse sky emission, obtained using likelihood analysis at lower energies by \citet{fermi_background}, is shown as a red shaded region. 
Pink upper limits above 100~GeV are from  \citet{ackerman_texas}. Black data points show the total point-source-subtracted flux from the North and South Galactic pole regions at $|b|\ge 60^\circ$. Solid line errorbars show  statistical error. Dashed line errorbar show the systematic error at the level of $\simeq 20$\% stemming from the uncertainty of the the Instrument Response Functions (IRF) of LAT (see {\tt http://fermi.gsfc.nasa.gov/ssc/data/analysis/LAT\_caveats.html}). Black horizontally shaded region shows the point source subtracted flux in the Galactic Pole regions found by \citet{fermi_background}. Grey data points and grey curve show the estimate of the residual cosmic ray background in the event set used in this analysis. The residual cosmic ray background level in the data set considered by \citet{fermi_background} is shown by the grey shaded region. Blue shading shows the Galactic diffuse emission in the North/South Galactic 
Pole regions $|b|>60^\circ$ derived by \citet{fermi_background}. Blue line shows the  Galactic diffuse background spectrum.}
\label{fig:spectrum}
\end{figure}

EGB flux could be found by subtracting the point source, Galactic diffuse and residual cosmic ray contributions to the overall number of events at high Galactic latitudes in each energy bin. The  spectrum of EGB obtained in this way is shown by the red thick data points in Fig. \ref{fig:spectrum}. The error in the measurement of the EGB flux at the energies below 100~GeV has contributions from the uncertainty of the level of the residual cosmic ray background as well as from the systematic uncertainties of the Instrument Response Functions (IRF) and of the Galactic diffuse background in the relevant sky region. An additional contribution to the error in the VHE band is given by  the statistical error arising from the low signal statistics. Finally, one more uncertainty stems from the uncertainty of the shape of the cosmic ray electron spectrum in the TeV energy range, which propagates to the uncertainty of extrapolation of the inverse Compton component of the Galactic diffuse emission above 100 GeV.

In the same figure we compare the measurement of EGB obtained from the direct photon counting at high Galactic latitudes with the results of the likelihood analysis of the all-sky data by \citet{fermi_background}. The two measurements agree well. 

Pink arrows at the energies above 100~GeV show the upper limit on the VHE EGB derived by \citep{ackerman_texas} using the likelihood analysis of the all-sky data. These upper limits agree with the upper limits derived from the direct photon counting (shown by the red arrows in Fig. \ref{fig:spectrum}. 

As a matter of fact, the level of Galactic diffuse emission at high Galactic latitudes turns out to be comparable to the level of EGB in the entire energy range $E>10$~GeV.\footnote{There is no a-priori reason why the two fluxes should be nearly equal. Thus, the equality of the two contributions poses a "fine-tuning" problem which requires further investigation. }
The Galactic diffuse emission contribution to the total flux at high Galactic latitude could not be negligibly small in the 100~GeV band. Taking this into account, it is not surprising that our estimate of the VHE EGB flux is  somewhat lower than the total diffuse emission flux at high Galactic latitudes and is, respectively, lower than the upper limit derived by \citet{ackerman_texas}. 

It is useful to note that extrapolating the EGB spectrum as a powerlaw spectrum to $E\ge 100$~GeV band would give the spectrum consistent with the data above 100~GeV.  With the current LAT exposure, there is still no evidence for suppression of VHE EGB flux due to absorption on EBL. A larger exposure time is needed to verify the presence of the feature. As it is mentioned in the Introduction, suppression of the flux above 100~GeV due to the absorption of VHE \gr s on the EBL is expected if the EGB is accumulated over the  cosmological distance scale. Detection of such suppression would be an important test of the origin of EGB.  

\section{$E>30$~GeV Extragalactic \gr\ background from point sources}

 As it is mentioned in the Introduction, different types of point and diffuse sources could contribute to the EGB in the VHE band. The main class of extragalactic point sources detected by {\it Fermi} is blazars, which are divided onto two sub-classed: BL Lac type objects and Flat Spectrum Radio Quasars (FSRQ). Over the first year of operation LAT has detected some $\sim 700$ such objects above 100~MeV energy \citep{fermi_agn}. BL Lacs and FSRQ have somewhat different spectral characteristics in the \gr\ band, with the spectra of BL Lacs being systematically harder than the spectra of FSRQ \citep{fermi_agn}. Hardness of the spectra of BL Lacs implies that they might produce significant contribution to the overall \gr\ flux in the VHE band. In fact, most of the extragalactic VHE \gr\ sources detected up to now by the ground based \gr\ telescopes sensitive above 100~GeV are BL Lacs\footnote{For the catalogs of extragalactic VHE \gr\ sources see e.g. {\tt http://tevcat.uchicago.edu} and {\tt http://www.isdc.unige.ch/vhe/index.html}}. 

\begin{figure}
\includegraphics[height=\columnwidth]{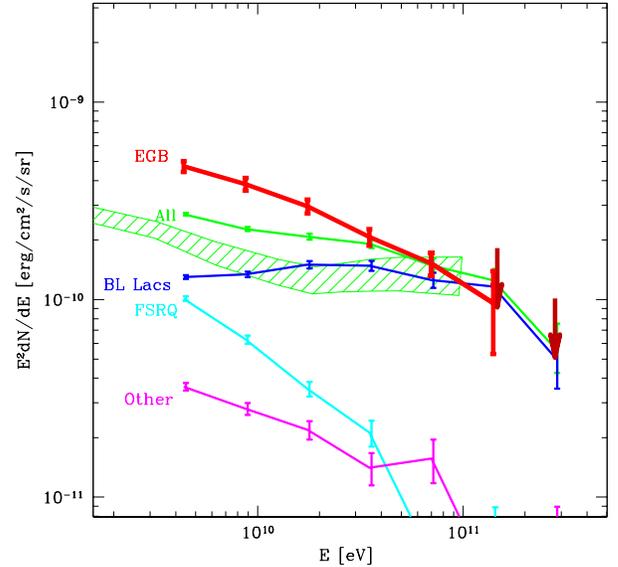}
\caption{Cumulative \gr\ flux from different classes of resolved  point sources. Source class is marked to the left from each curve. Green shaded area shows the point source flux at high Galactic latitudes found by \citet{fermi_background}. Red curve shows the EGB spectrum from Fig. \ref{fig:spectrum}, which includes only unresolved sources. }
\label{fig:sources}
\end{figure}

Fig. \ref{fig:sources} shows the breakdown of the point source contributions to the high Galactic latitude flux by the source type. One could clearly see that the dominant contribution is given by BL Lac objects which provide $\ge 90\%$ of the total point source flux above $30$~GeV. The cumulative spectrum of the other major blazar class, FSRQ has a high-energy cut-off at $\sim 10$~GeV so that FSRQ contribution to the point source flux is negligible in the VHE band. From Fig. \ref{fig:sources} one could see that the total point source flux calculated from the cumulative photon distribution around stacked point sources in the high Galactic latitude regions (see Section \ref{sec:ps}) is in a good agreement with the total point source flux calculated using the likelihood analysis by \citet{fermi_background}, shown by the green shaded region.
Above 50 GeV  $ 90\%$  of source photons  come from BL Lacs and $10\% $ from "Other" Fermi sources, which are dominated by not-identified sources with some contribution from nearby AGN's. 

The EGB spectral shape above 30~GeV  follows the cumulative point source spectrum. This observation leads to a conjecture that the VHE EGB is produced by already known type of VHE \gr\ point sources with fluxes below the sensitivity of LAT. Since the dominant source class in the VHE band is that of BL Lacs, a more precise conjecture is that the VHE EGB is produced by the unresolved BL Lacs. 

\section{BL Lac contribution to VHE EGB}

In the unification schemes of AGN BL Lac objects are identified with the Fanaroff-Riley type I  (FR I) radio galaxies with jets aligned with the line of sight \citep{urry95}. This implies that the cosmological evolution of the BL Lacs should follow that of the FR I radio galaxies. Recent studies of the cosmological evolution of  FR I galaxies show that they experience "positive" cosmological evolution, which is usually described in terms of  luminosity or comoving source density evolution as the increase of either average source luminosity or the average comoving source density with the redshift $z$, as $(1+z)^k,\ \ k>0$. Different recent studies find somewhat different values of $k$, depending on the analyzed radio galaxy samples and different assumptions about the evolution type (luminosity or density), with $k$ ranging in $1\lesssim k\lesssim 3$ \citep{sadler07,hodge09,smolcic09}. Since BL Lacs are just the FR I galaxies specially oriented with respect to the line of sight, their cosmological evolution follows the evolution of FR I galaxies, with the increasing source luminosity or spatial density with the redshift. This implies that significant flux should be produced by the sources at large redshifts, $z\sim 1$. It is possible that most of  individual sources at large redshifts are too weak to be significantly detected by LAT, but collective emission from all the set of BL Lacs at high redshifts  gives a significant contribution to the EGB.

\begin{figure}
\includegraphics[angle=-90,width=\columnwidth]{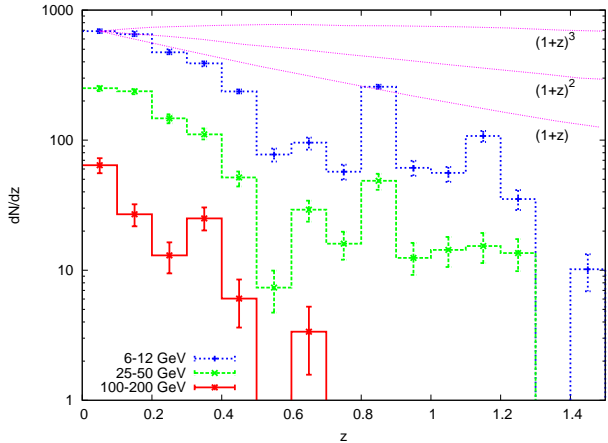}
\caption{Number of detected VHE photons as a function of redshift in the 6.25-12.5~GeV (blue dotted histogram) 25-50~GeV (green dashed histogram) and 100-200~GeV (red solid histogram) bands. Also shown are the distributions of photons with the redshift expected for different laws of BL Lac cosmological evolution. }
\label{fig:dNz}
\end{figure}

Dependence of the total flux of the BL Lac population on the redshift could be found from the following straightforward calculation. 
Let us consider the total flux produced by sources at redshift $z$ in a redshift interval $\Delta z$. This redshift interval corresponds to the comoving distance interval $\Delta r=\Delta z/H(z)$ where $H(z)\sim\sqrt{\Omega_\Lambda+\Omega_m(1+z)^3}$ is the expansion rate of the Universe filled with matter and cosmological constant with today's densities $\Omega_m$ and $\Omega_\Lambda$. 

As an example, we take the case of "pure luminosity" evolution with the average source luminosity increasing as $(1+z)^k$ and conserved comoving source density $n(z)=n_0=const$. The number of sources in a spherical layer of thickness $\Delta z$ is $\Delta N_s=4\pi n_0 r^2\Delta r$. Each source produces the flux in a given energy band  $F\sim (1+z)^{k-\Gamma}/(4\pi r^2)$, where $\Gamma$ is the photon index, the factor $(1+z)^{1-\Gamma}$ describes the change in the number of photons in a given energy band due to the cosmological redshift of the photon energies. One power of $(1+z)$ is compensated by the time delay between subsequent photons. 

The flux from the sources at large redshifts is affected by absorption of VHE photons on EBL. For example, at $z\simeq 1.5$ the absorption modifies source spectrum above the energy $E\simeq 50$ GeV, if one assumes the EBL evolution calculated by \citet{franceschini08}. Absorption on EBL leads to suppression of the flux by a factor $\exp\left(-\tau(E,z)\right)$ where 
 $\tau(E,z)$ is the optical depth with respect to the pair production.
 
The overall flux from the sources in the redshift interval $\Delta z$ is 
\begin{equation}
\frac{\Delta F(E,z)}{\Delta z}=F\Delta N_s\sim \frac{(1+z)^{k-\Gamma}e^{-\tau(E,z)}}{\sqrt{\Omega_\Lambda+\Omega_m(1+z)^3}}
\label{dfdz}
\end{equation}

Fig. \ref{fig:dNz} shows the number of \gr s as a function of source redshift. For this we used BL Lacs with known redshifts  from Veron\&Veron \citep{veron13} catalog complemented by BL Lacs detected by LAT, but not listed in the Veron\&Veron catalog. Only sources with $|b|>10^\circ$ were considered. 
Here we plot photon distributions from Fermi BL Lacs  in the  three energy bands: $6.25-12.5$ GeV, $25-50$ GeV and $100-200$ GeV.   One can see that at lower energies $E<50$ GeV a  significant flux is produced by BL Lacs at large redshifts  up to  $z=1.5$. At the highest energies only contribution from nearby sources at $z<0.7$ is present. Two effects might explain the deficit of high-redshift sources at high energies. First, the flux at the highest energies is suppressed by absorption on EBL. Next, the photon statistics in the highest energy bin is low so that  sources contributing to the flux in the 6.25-12.5~GeV bin produce less than one photon in the 100-200~GeV bin. 

In the same figure we also show the expected dependence of the number of photons on the redshift expected in different  evolution models,  Eq.~(\ref{dfdz}). The models for cases $k=1,2,3$ are shown with magenta lines for $6.25-12.5$ GeV energy band. We normalize the models to the number of photons in the first redshift bin, in which we have the most complete knowledge of the BL Lac population. 

From the comparison of the evolution models with the data one might get an impression that $(1+z)$ model is more consistent with the data than the models assuming faster evolution. However, the histogram on Fig. \ref{fig:dNz} does not take into account photons from BL Lacs with unknown redshifts. These BL Lacs produce  about 30 \% of all cumulative BL Lac flux. This means that at least 30\% contribution to the overall flux (integrated over all redshifts) is missing in  Fig. \ref{fig:dNz}.  
The model with evolution $k=1$ predicts the total number of photons which is $\sim 3\sigma$  below the total number of photons in BL Lacs with known and unknown redshift together in the $6.25-12.5$ GeV energy band. In the energy band 3.125-6.25~GeV the under-prediction of the total number of photons from BL Lacs in the  $k=1$ model  is at  $\ge 5\sigma$  level, which means that the model is efficiently ruled out. Thus, Fermi LAT observations of BL Lac objects indicate that BL Lacs have positive cosmological evolution with $k>1$.

In all other cases $k>1$, the discrepancy between the observed and expected number of photons from BL Lacs starts already at small redshifts, $z\ge 0.2$. The "missing BL Lac" \gr s could come either form BL Lacs with unknown redshifts or from {\it Fermi} sources which are not yet identified as BL Lacs or, finally, from BL Lacs with fluxes below the sensitivity of LAT.

\begin{figure}
\includegraphics[height=\columnwidth]{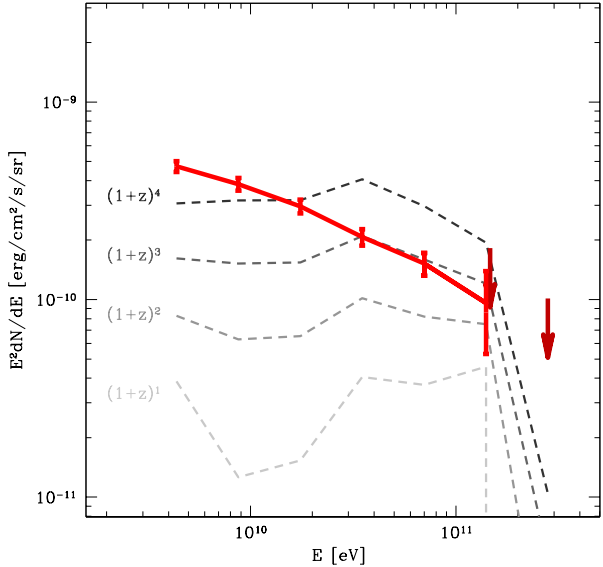}
\caption{VHE EGB produced by unresolved BL Lacs under different assumptions about the cosmological evolution of BL Lac population (the evolution law is marked to the left of each curve). Red data points show the EGB spectrum from Fig. \ref{fig:spectrum}.}
\label{fig:evolution}
\end{figure}

Although individual high redshift BL Lacs would not be detectable by LAT, cumulative flux of these BL Lacs could give significant contribution to the EGB. Fig. \ref{fig:evolution} shows the contributions from "missing BL Lacs" at high redshifts up to $z=1$ expected in four different models of cosmological evolution of BL Lac / FR I population. To calculate this contribution, we have normalized $\Delta F/\Delta z$ distribution on the measured flux of BL Lacs in the redshift bin $0<z<0.1$. For the two last bins,100-200~GeV and  200-400~GeV, the statistics of the BL Lac signal is too low to normalize $\Delta F/\Delta z$ by the flux in the first redshift bin. To estimate the normalization of $\Delta F/\Delta z$ in these energy bins we have assumed that the average BL Lac spectrum extends as a powerlaw with photon index $-2$ up to the 400~GeV energy range. The resulting statistics of the signal in the last 200-400 bin in Fig. \ref{fig:evolution} is low and is subject to large fluctuations.

From Fig. \ref{fig:evolution} one could see that the conjecture that EGB is produced by collective emission from distant BL Lacs is valid if BL Lacs experience positive luminosity or density evolution of the form  $(1+z)^k$ with $k\simeq 3$. Slower or faster cosmological evolution with $k=2$ or $k=4$, respectively, would under- or over-produce the EGB.  Since evolution $k = 4$ predicts number of photons more then
 diffuse gamma-ray background (see Fig. \ref{fig:evolution})  it is also excluded. Thus, the LAT data impose a constraint on the cosmological evolution of the BL Lac/ FR I population 
 \begin{equation}
 1< k < 4
 \end{equation} 
From Fig. \ref{fig:evolution} one could also see that if $k=3$, all the flux of EGB above 30~GeV could be explained by a cumulative emission from the BL Lac population. At the same time, if $k$ is significantly smaller than 3, as it is suggested by some recent studies of the evolution of the parent population of FR I galaxies \citep{smolcic09}, significant part of the VHE EGB flux should come from a yet unknown source population or have a truly diffuse nature.  

\gr\ flux from high-redshift BL Lacs is modified in the VHE band by the effect of absorption on EBL. The model spectra shown in Fig. \ref{fig:evolution} take this effect into account. We use the model of \citet{franceschini08} to estimate the attenuation of the VHE \gr\ flux from redshifts up to $z=1.5$. Model curves shown in Fig. \ref{fig:evolution} assume that the average intrinsic spectrum of BL Lacs does not have a high-energy cut-off up to $\sim 400$~GeV. The observed suppression of the flux above 100~GeV is explained only by the effect of absorption on EBL.

\section{Discussion and Conclusions}

In this paper we have derived a measurement of  EGB in the 10-400~GeV energy range from the analysis of {\it Fermi}/LAT data in the North and South Galactic Poles regions, $|b|>60^\circ$. Our approach was to count all the protons  detected by LAT in this region and estimate the number of counts from the point sources, from the Galactic diffuse emission, form the residual cosmic ray background and from the EGB. Subtracting the source, Galactic diffuse and residual cosmic ray background counts from the total number of counts in the North and South Galactic Pole regions we derived the EGB spectrum shown in Fig. \ref{fig:spectrum}.

Comparing the spectrum of EGB in the $>30$~GeV energy band with the  spectrum of extragalactic point sources in the same energy band (Fig. \ref{fig:sources}), we have noticed that the two spectra closely follow each other. Based on this observation, we have put forward a conjecture that the EGB above 30~GeV is explained by the unresolved BL Lacs, which give dominant contribution to the extragalactic point source flux in this energy band. We have demonstrated that this conjecture is consistent with the EGB measurement provided that BL Lacs follow positive cosmological evolution with the overall power of emission from the source population increasing as  $(1+z)^3$ up to $z\sim 1$ (Fig. \ref{fig:evolution}). 
Such cosmological evolution is roughly consistent with the measurements of cosmological evolution of FR I radio galaxies which are believed to be the parent population of BL Lac type objects and are also observed to have positive cosmological evolution of the form $(1+z)^k$ with an uncertain value of $k$ between 1 and 3  \citep{rigby08,smolcic09,sadler07}. At the same time, it is opposite to the negative cosmological evolution of the high-energy-peaked BL Lacs \citep{giommi99,giommi01}, which constitue a sub-class of the GeV-TeV \gr\ emitting BL Lacs considered in our analysis. 

If the VHE EGB is indeed produced by distant BL Lacs at redshifts up to $z\sim 1$, LAT will not be able to resolve it into point sources. Indeed, the brightest BL Lac on the sky, Mrk 421 produced $\simeq 30$ photons above 100~GeV. If the positive cosmological evolution of BL Lac population is mostly due to the increase of the comoving source density rather than increase of the typical source luminosity, the brightest BL Lacs at redshift $z\sim 1$ produce $\sim 10^{-2}$ \gr s in LAT over some 2.5 years of exposure. This means that LAT would not collect sufficient photon statistics to detect distant BL Lacs individually. If the $(1+z)^3$ evolution is mostly due to the increase of the average source luminosity, BL Lacs at redshift 1 have an order-of-magnitude higher luminosity than local BL Lacs. However, even with higher luminosity, they produce  on average 0.1 photon in LAT, so that they are still not individually detectable.

If the real value of $k$ is much below $k=3$, as indicated by a recent study by \citet{smolcic09}, emission from the unresolved BL Lacs would not explain the VHE EGB flux and there should be another source class or a mechanism of production of diffuse emission which would account for the EGB. Such a mechanism might, in fact, be indirectly related to the BL Lac population. Most of the power output from BL Lacs at the energies above 100~GeV is converted into the electromagnetic emission from \gr\ induced cascade in intergalactic medium.  The intrinsic spectra of BL Lacs (and of FR I radio galaxies, such as M87 and Cen A \citep{m87,cena}) are known extend up to  $\sim 10$~TeV. Typical energy of the cascade photons which are produced via inverse Compton scattering of CMB photons by the $e^+e^-$ pairs deposited in the intergalactic medium is $E_\gamma\simeq 100\left[E_{\gamma_0}/10\mbox{ TeV}\right]^2\mbox{ GeV}$ \citep{neronov09}. If the intrinsic source luminosities in the 1-10~TeV range are comparable to the luminosities in the 10-100~GeV, total flux of the cascade emission in the 10-100~GeV band is expected to be comparable to the point source flux, so that the cascade emission could give significant contribution to the EGB \citep{coppi97}. This is consistent with the observation that the VHE EGB level is comparable to the cumulative extragalactic  point source flux in the 10-100~GeV band observed by LAT.

The only possibility to test the hypothesis of BL Lac origin of EGB would be to use deep observations with ground-based \gr\ telescopes. Ground based \gr\ telescopes, which are sensitive in the VHE energy band, have much larger collection area above several hundreds of GeV and, as a consequence, could detect much weaker sources, than LAT. The flux from Mrk 421-like BL Lacs at the redshift $z\simeq 1$ is $\sim 10^{-13}$~erg/cm$^2$s. If the cosmological evolution of BL Lacs is due to increase of the average source luminosity with redshift, brightest BL Lacs at redshift $z\sim 1$ might produce fluxes up to $10^{-12}$~erg/cm$^2$s at 100~GeV. At the energies around $E\lesssim 100$~GeV this flux is not strongly attenuated by the absorption on EBL. 
The energy $E\simeq 100$~GeV is around or below the low energy threshold of the  current generation Cherenkov telescopes, like HESS, MAGIC and VERITAS. However, next generation facilities, Cherenkov Telescope Array (CTA) \citep{cta} or 5@5  \citep{5at5} are expected to have an energy threshold significantly below 100~GeV. Their sensitivity could be sufficient to  resolve the VHE EGB into point sources, at least in the case when the cosmological evolution is mostly luminosity, rather than source density evolution.

 \section*{Acknowledgements}
 
 We would like to thank I.Moskalenko, for the discussion of the issues related to the Galactic \gr\ background, and M.Ackermann for the clarification of the uncertainties of the residual cosmic ray background. The work of AN is supported by the Swiss National Science Foundation grant PP00P2\_123426.

\end{document}